\documentclass[aps,twocolumn,groupedaddress,showpacs]{revtex4}
\usepackage{graphicx}

\begin{document}
\bibliographystyle{apsrev}

\title{Doping evolution of the phonon density of states and electron-lattice interaction in  
Nd$_{2-x}$Ce$_x$CuO$_{4+\delta}$\\}
\author{H. J. Kang,$^1$ Pengcheng Dai,$^{1,2,\ast}$ D. Mandrus,$^{2,1}$ R. Jin,$^2$ H. A. Mook,$^2$
D. T. Adroja,$^3$ S. M. Bennington,$^3$ S.-H. Lee,$^4$ and J. W. Lynn$^4$  
 }
\address{$^1$Department of Physics and Astronomy, The University of Tennessee, Knoxville, Tennessee 37996-1200}
\address{$^2$Solid State Division, Oak Ridge National Laboratory, Oak Ridge, Tennessee 37831}
\address{$^3$ISIS facility, Rutherford Appleton Laboratory, Chilton, Didcot OX11 0QX, UK}
\address{$^4$NIST Center for Neutron Research, National Institute of Standards and Technology, 
Gaithersburg, Maryland 20899}
\date{\today}
\begin{abstract}
We use inelastic neutron scattering to study the evolution of the generalized phonon density of states (GDOS)
of the $n$-type high-$T_c$ superconductor Nd$_{2-x}$Ce$_x$CuO$_{4+\delta}$ (NCCO),
from the half-filled Mott-insulator ($x=0$) to the $T_c=24$ K superconductor ($x=0.15$). 
Upon doping the CuO$_2$ planes in Nd$_2$CuO$_{4+\delta}$ (NCO) with electrons by Ce
substitution, the most significant change in the GDOS is the softening of the highest 
phonon branches associated with the Cu-O bond stretching and out-of-plane oxygen vibration modes. However,
the softening occurs within the first few percent of Ce-doping and is not related to the electron 
doping induced nonsuperconducting-superconducting transition (NST) at $x\approx 0.12$. These results suggest that the electron-lattice coupling in the 
$n$-type high-$T_c$ superconductors is different from that in the $p$-type materials.
\end{abstract}

\maketitle

\narrowtext
One of the most remarkable properties of high-transition-temperature (high-$T_c$) copper-oxide (cuprate)
superconductors is their close proximity to an antiferromagnetic (AF) phase. The parent compounds of the  
high-$T_c$
cuprates are AF insulators characterized by a simple doubling of the crystallographic unit cell in the CuO$_2$ planes \cite{kastner}. When holes \cite{bednorz} or electrons \cite{tokura} 
are doped into these planes, the long-range AF-ordered phase is destroyed, and the copper-oxide materials become metallic and superconducting with persistent short-range AF spin correlations (fluctuations). 
Much effort over the past decade has focused on understanding the 
nature of the interplay between magnetism and   
 superconductivity \cite{kastner}, mainly because spin fluctuations may contribute 
a major part of the superconducting condensation energy \cite{demler,dai}. On the other hand,
the role of phonons  in the microscopic mechanism of superconductivity
 is still largely unknown even though phonons in cuprates 
also display a variety of unusual properties 
 \cite{pintschovius,mcqueeney1,mcqueeney2,pintschovius1,mook}. 
The key question is whether magnetism and electron-electron correlations alone are sufficient to induce electron pairing that leads to superconductivity in high-$T_c$ cuprates, or electron-lattice coupling 
also plays an important role. 

From the analysis of high-resolution angle-resolved photoemission (ARPES) data in conjunction with those from neutron, optics and local structural probes, Shen and co-workers \cite{shen} suggest that phonons must 
also play an essential role in electron pairing for high-$T_c$ cuprates. 
The key evidence for electron-lattice coupling, they argue \cite{lanzara}, is that 
the kink (or the change of slope) seen in the electronic 
dispersion of the hole-doped ($p$-type) Bi$_2$Sr$_2$CaCu$_2$O$_8$ (Bi2212), Bi$_2$Sr$_2$CuO$_6$ (Bi2201), 
and La$_{2-x}$Sr$_x$CuO$_4$ (LSCO) from the ARPES data \cite{valla,bogdanov,kaminski,johnson}
occurs at an energy ($\sim$70 meV) very close to the phonon anomalies observed by inelastic neutron
scattering \cite{mcqueeney1,mcqueeney2,pintschovius1}. 
These phonon anomalies include the break in the dispersion of the 
oxygen half-breathing mode in La$_{1.85}$Sr$_{0.15}$CuO$_4$ \cite{mcqueeney1} and the 
abrupt development of new oxygen lattice vibrations near the doping-induced 
metal-insulator 
transition (MIT) in the generalized phonon density of states (GDOS) of LSCO \cite{mcqueeney2,note}. 
Since the change of slope in the electronic dispersion indicates a 
dramatic drop in the ``quasiparticle'' scattering rate \cite{shen},
their observation in hole-doped cuprate superconductors 
\cite{valla,bogdanov,kaminski,johnson} suggests a strong coupling between the quasiparticles
and a sharp collective spin or lattice mode. Although the neutron magnetic resonance \cite{dai} could be  
the collective spin mode coupled to the quasiparticles \cite{johnson,norman}, Shen {\it et al.} argue that 
electron-lattice interaction is ultimately responsible for the quasiparticle velocity change and thus is 
crucial to the high-$T_c$ superconductivity \cite{shen,lanzara}. Furthermore, since the 
dispersion of the electron-doped superconducting Nd$_{1.85}$Ce$_{0.15}$CuO$_{4+\delta}$ does not have such a kink, 
the authors \cite{shen} predict that the $n$-type materials have much weaker electron-lattice coupling and thus lower 
$T_c$'s. 

If this hypothesis were correct, one would expect the exotic lattice dynamics seen in the $p$-type
 LSCO  
\cite{mcqueeney1,mcqueeney2,pintschovius1} 
to be reduced in the $n$-type Nd$_{2-x}$Ce$_x$CuO$_{4+\delta}$ (NCCO) \cite{shen}.  For LSCO,  
the abrupt development of the new oxygen lattice vibrations 
across the doping induced nonsuperconducting-superconducting transition (NST)  
was interpreted as evidence for strong electron-lattice coupling in the 
superconducting cuprates that is not present in nonsuperconducting 
materials \cite{mcqueeney2,note}. Specifically, the new lattice mode 
at $\sim$70 meV in the GDOS is believed 
to be at least partly comprised 
of the anomalous Cu-O bond-stretching (oxygen half-breathing) mode \cite{mcqueeney2}. 
Although the GDOS for NCCO with $x=0, 0.15$ were studied by Lynn and co-workers \cite{lynn,sumarlin}, 
no systematic doping dependent measurements are available.
If the quasiparticle velocity drop seen in the ARPES data of $p$-type cuprates
\cite{shen} is related 
to the anomalous lattice vibrational modes \cite{mcqueeney2}, the absence of such a drop in the electron-doped 
superconducting NCCO with $x=0.15$ 
would suggest a weak (or no) phonon anomaly for NCCO.   

In this paper, we present inelastic neutron scattering measurements of the GDOS in 
NCCO spanning electron doping concentrations from the half-filled Mott-insulator 
(Nd$_2$CuO$_{4+\delta}$ or NCO) to optimally doped NCCO
superconductor ($x\approx 0.15$). Upon doping electrons to the CuO$_2$ planes 
by Ce substitution, the most significant change in the GDOS is the anomalous softening of the  
$\sim70$ meV phonon branches associated with the oxygen half-breathing and out-of-plane 
vibrational modes. However, in contrast to LSCO \cite{mcqueeney2},
the anomaly only occurs within the first few percent of Ce-doping  
and there is 
no evidence for any new lattice modes in the GDOS across the electron 
doping induced NST at $x\approx 0.12$. 
Our results indicate that doping-induced phonon anomalies in hole-doped cuprates are different from these
in electron-doped materials, thus suggesting that electron-lattice coupling is
unrelated to the superconductivity of NCCO.

Our experiments were performed on the MARI chopper spectrometer at ISIS facility, 
Rutherford Appleton Laboratory, UK. 
The detectors on MARI cover a wide scattering angle from 3$^\circ$ to 135$^\circ$. 
For the experiments, we used a Fermi chopper to choose an 
incident beam energy of 110 meV.
The energy resolution is between 1-2\% of the incident 
energy. The powder samples were mounted inside the aluminum sample can 
on the cold head of a helium closed-cycle refrigerator and all 
measurements were performed at $T=30$ K. The incident neutron beam size is $5\times 5$ cm$^2$ and
the unexposed area of the sample was covered by Cd sheets. 
To normalize the scattering from NCCO on an absolute scale, 
we used the elastic incoherent scattering from a vanadium standard. 
In addition to 
measurements at MARI, we have also collected data on the BT-4 filter analyzer spectrometer
at the National Institute of Standards and Technology research reactor. To 
within the error of the
measurements, the results of these two experiments are identical. 

\begin{figure}
\includegraphics[keepaspectratio=true, totalheight = 2.1 in, width = 2.1 in]{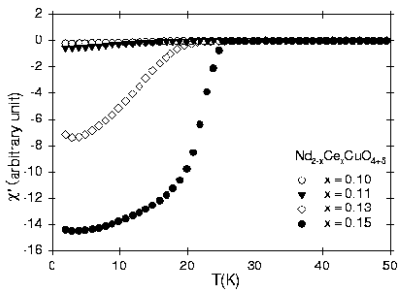}
\caption{
Temperature dependences of the AC magnetic susceptibility $\chi^\prime$ (real
part) for the NCCO powder samples used in
the neutron measurements. The background was subtracted using the
$\chi^\prime$ for $x=0.09$, which shows Curie-Weiss-like behavior down to the lowest
temperature measured. The diamagnetic signal first appears for $x\geq 0.11$.
}
\end{figure}

We prepared the ceramic samples of NCCO with Ce-concentrations of $x=0.00$, 0.04, 0.08,
 0.09, 0.10, 0.11, 0.12, 0.13, 0.15 by the conventional solid state reaction \cite{tokura}. 
The as-grown samples have excess oxygen ($\delta>0$) and are nonsuperconducting. 
Various annealing procedures have been developed to remove the
excess oxygen needed to produce superconductivity.  However, the properties of the samples and the resulting electronic phase diagrams are different depending on the details of the annealing procedure used.  
In the original work of Tagaki {\it et al.} \cite{tokura}, it was found that samples treated in flowing Ar at temperatures in excess of 1100 $^\circ$C followed by heating in air at 500 $^\circ$C produced metallic samples with a sharp superconducting transition.  However, it was also found that this procedure resulted in some decomposition of the sample as well as probable loss of Cu from the surfaces of the polycrystalline 
grains \cite{tokura}. This procedure also results in the electronic phase diagram showing an abrupt
NST around $x=0.14$ with only half of the superconducting ``dome'' \cite{tokura}, different from 
hole-doped LSCO \cite{kastner}. We have followed the annealing procedure developed by Maple's group \cite{maple}, 
where the samples are treated in flowing Ar at temperature of about 900 $^\circ$C. The resulting phase diagram 
shows the NST around $x=0.12$ with the almost complete superconducting dome \cite{maple} as compared to the half-dome 
from \cite{tokura}. To characterize the materials, bulk magnetization and resistivity
measurements were performed for all the samples. Figure \ 1 shows the doping dependence of
the AC susceptibility for $x=0.1$, 0.11, 0.13, and 0.15.  Superconductivity is clearly
seen for NCCO with $x\geq0.13$, thus confirming that the NST in NCCO occurs around $x\approx 0.12$ \cite{maple}.

In an unpolarized neutron experiment, the major difficulty 
in obtaining the reliable GDOS is to separate phonons from magnetic scattering.
For NCCO, the largest magnetic signal  
originates from single-ion crystalline electric field (CEF) 
excitations of the Nd ions \cite{lynn,sumarlin}.
The CEF excitations 
of NCCO with $x=0$, and 0.15 have been studied in great detail 
and their level scheme has
peaks around $\hbar\omega \approx$ 12--16, 
$20.5$, $27$, and $93.3$ meV at low temperatures \cite{boothroyd}.
We performed careful wave vector ($Q$) dependent analysis of the  
excitation intensities at $\hbar\omega =$ 20.5, 27, and 93.3 meV for 2 \AA $^{-1}<Q<10$ \AA $^{-1}$. 
The outcome  
confirms the earlier results that these three peaks are magnetic in origin and the phonon cutoff 
energy of NCCO is around $83$ meV \cite{sumarlin}. 
We also checked the strength of
the multiple scattering and multi-phonon scattering using the Monte Carlo simulation program MSCAT,
but found such multi-phonon scattering contributes negligibly
to the total scattering intensity 
in the energy region of interest ($\hbar\omega\geq 50$ meV).
To reduce the magnetic scattering
contribution to the GDOS, we replaced the  
intensities of the 20.5 and 27 meV
peaks with scattering from the highest measured wave vectors (9 \AA $^{-1}<Q<11$ \AA $^{-1}$).
Although this procedure may not eliminate all the magnetic intensity,
there are no magnetic contributions to the GDOS for 50 meV$\leq\hbar\omega\leq80$ meV.

\begin{figure}
\includegraphics[keepaspectratio=true, width=0.75\columnwidth,clip]{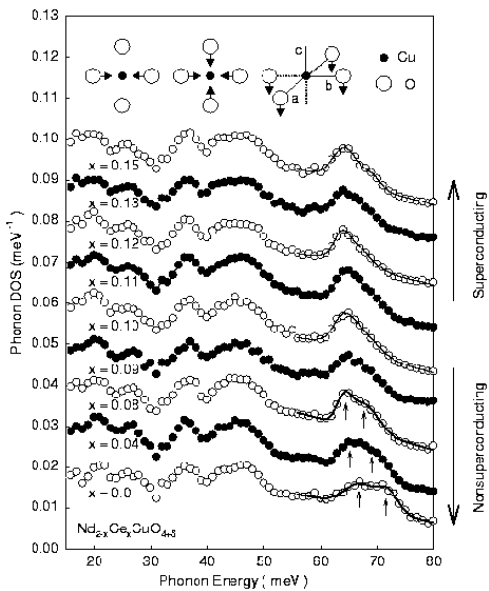}
\caption{
The GDOS of NCCO as a function of $x$ at $T=30$ K. Each GDOS is displaced along
the vertical axis for clarity and the solid lines are Gaussian fits discussed in the text.
The nonsuperconducting-superconducting transition as a function of $x$ is schematically shown on the
right. The insets show the polarizations of the oxygen half-breathing (left), breathing (middle), 
and out-of-plane (right) vibrational modes.
}
\end{figure}

After subtraction of the empty aluminum sample can, multiple and multi-phonon
scattering, single-phonon GDOS with 
$Q$ values integrated from 3 to 11 \AA$^{-1}$ 
were calculated by multiplying
$\omega/[n(\omega)+1]$, where $n(\omega)$ is the Bose population factor. The total area of each
GDOS was then normalized to 1 over the energy range from 15 to 80 meV. Figure 2 shows the 
GDOS for NCCO with $x=0.0$, 0.04, 0.08, 0.10, 0.11, 0.12, 0.13, and 0.15. Consistent with previous measurements 
on NCCO for $x=0.0$ and 0.15 \cite{lynn,sumarlin}, the spectra contain clear 
peaks at $\sim$36, 42, 48, and 65-70 meV. On moving from an insulator to a metal with 
increasing Ce-concentration, 
the largest observed effect is the softening and sharpening of the broad $\sim$70 meV phonon-band
in the undoped NCO. 

We systematically fit the 70 meV phonon band 
with two Gaussians on a sloping background for various $x$.
The solid lines in Fig. 2 show the outcome of the fits.   
Although the precise functional form of the GDOS for the 70 meV phonon band is not known, 
the systematic Gaussian fits 
allow a quantitative determination for the magnitude of the phonon softening.
For the undoped NCO, the 70 meV mode shows a flattish top and 
can be best fitted by two Gaussians centered at 67 and 71 meV,
respectively. On increasing the Ce-concentration to $x=0.04$, 
the 71 meV mode softens to 67 meV (6\% softening) 
and shows less of a flattish top. 
Furthermore, the GDOS gains intensity at 65 meV at the expense of the 71 meV peak. 
At $x=0.08$, the GDOS peaks more sharply at 65 meV. On further increasing 
$x$ and across the NST at $x=0.12$, the GDOS show essentially no change
from that for $x=0.08$ to within the error of the measurements. 
Figure 3 shows the comparison plots of the GDOS at various
$x$. For NCCO with $0.0\leq x\leq 0.08$, the 70 meV phonon band shows 
significant softening while all other modes display no visible change with 
increasing $x$ (Figs. 3a and 3b).
The GDOS for $x$ from 0.08 to 0.15 (Figs. 3c and 3d)
overlap completely in the probed energy range ($15\leq \hbar\omega \leq 80$ meV) 
and show no changes across the NST.

\begin{figure}
\includegraphics[keepaspectratio=true, width=0.65\columnwidth,clip]{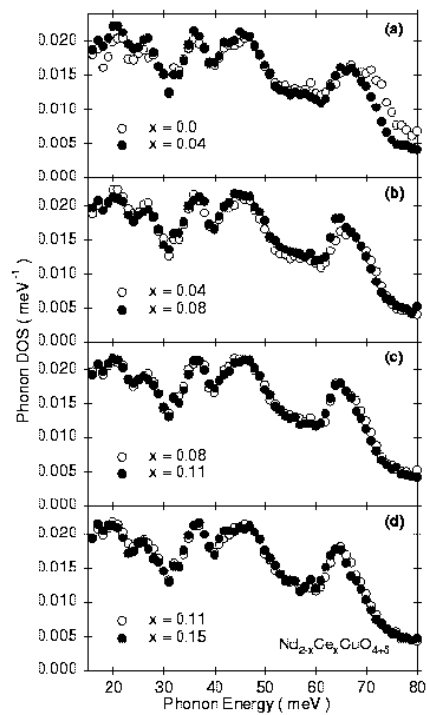}
\caption{
Comparison of the GDOS of NCCO as a function of electron-doping, $x$. (a) The GDOS of NCCO for 
$x=0.0$, and 0.04; (b) $x=0.04$, and 0.08; (c) $x=0.08$, and 0.11; and (d) $x=0.11$, and 0.15.
}
\end{figure}

To understand the atomic 
displacement patterns of the phonon modes contributing to the 70 meV band in NCO, 
we consider its experimentally determined phonon dispersion curves \cite{pintschovius2}. 
For NCO, the highest energy phonon bands  
are around 70 meV \cite{pintschovius2}. These include the highest energy 
in-plane Cu-O bond-stretching mode with
$\Delta_1$-symmetry at $Q=(0.5,0,0)$ (the oxygen half-breathing mode) along the $[\zeta,0,0]$ direction, the 
out-of-plane ($c$-axis polarized) oxygen breathing mode with 
$\Lambda_1$-symmetry along
$[0,0,\zeta]$, and the in-plane oxygen breathing mode with 
$\Sigma_1$-symmetry at $Q=(0.5,0.5,0)$ along
$[\zeta,\zeta,0]$.   
The inset of Fig. 2 shows the oxygen displacement patterns for these three modes. 
Since these three modes are at $\sim$70 meV in the dispersion curves \cite{pintschovius2},
the 70 meV peak in the GDOS of NCO must consist, at least partially, 
of these modes. As a consequence, the electron-doping induced softening 
in NCCO must also occur in these modes.

In a very recent inelastic X-ray scattering study of longitudinal optical phonons in 
NCCO with $x=0.14$, d'Astuto {\it et al.} discovered anomalous phonon softening 
in the two highest longitudinal branches associated with the Cu-O bond-stretching 
and out-of-plane oxygen vibrations \cite{dastuto}. By comparing their data on
NCCO with undoped NCO, the authors concluded that strong electron-phonon coupling is also present 
in electron-doped NCCO. 
From their work \cite{dastuto}, it becomes clear that the significant 
softening of the 70 meV phonon band with $x$ in Figs. 2 and
3 is mostly due to the softening of the oxygen half-breathing and out-of-plane vibrational 
modes. For hole-doped LSCO, the 
oxygen half-breathing modes display anomalous 
behavior \cite{mcqueeney1,pintschovius1} and show up as new lattice modes 
in the superconducting side of the phase diagram across the NST \cite{mcqueeney2}. While 
the oxygen half-breathing modes also exhibit anomalous behavior \cite{dastuto} and soften 
with increasing electron-doping, our results indicate that the softening occurs  
within the first few percent of Ce-doping in the nonsuperconducting regime and therefore 
is not associated with the electron doping induced NST in NCCO. 

For $p$-type cuprates, previous investigations 
have established a clear correlation between superconducting properties of the 
materials and special features of the phonon spectrum. While such correlation is seen 
as anomalous phonon modes across the NST in LSCO \cite{mcqueeney2}, 
systematic studies of the GDOS in YBa$_2$Cu$_3$O$_{6+x}$ show that 
the phonon cut-off energy softens across the NST 
and is closely related to $T_c$ (see Fig. 41 of \cite{pintschovius}). In general, 
these phonon anomalies are related to the dielectric screening properties of metals 
and thus suggest a strong electron-lattice coupling in the superconductivity of the $p$-type  
materials. Although phonon softening is also observed in the $n$-type 
NCCO, our data indicate that these anomalies occur in the nonsuperconducting regime and 
are not directly related to the NST. Therefore, it becomes clear that  
the electron-lattice coupling in the electron-doped NCCO is different from 
that in the hole-doped materials. 

It is interesting to compare our results with that of the ARPES on NCCO. 
In principle, the strong electron-lattice coupling and large softening of 
the optical oxygen vibrational modes in NCCO with $x=0.04$ 
should reveal themselves as distinctive features in the ARPES spectra \cite{shen}. If
the kink in the electronic band dispersion in the hole-doped materials 
is due to the anomalous softening of 
the 70 meV oxygen half-breathing mode \cite{shen}, its absence in electron-doped 
NCCO would suggest no softening of such oxygen modes in NCCO. Clearly,
this is inconsistent with the results of \cite{dastuto} and the present work. On the 
other hand, if the kink in the ARPES spectra is not related to the softening of
the 70 meV modes but to the changes 
of such modes across the hole- (electron-) induced NST, our data would be consistent 
with a weak electron-lattice coupling in NCCO. 
An unambiguous test of this idea will require comparison of the neutron data with 
the doping dependence of the electronic structure of NCCO. Although systematic 
ARPES investigations have been carried out very recently on NCCO \cite{armitage}, 
the evolution of the electronic dispersions across the MIT or NST is unavailable and
therefore cannot be compared yet with the neutron results. 
 
We thank Z.-X. Shen and Y. Ando for helpful discussions. 
This work was supported by NSF DMR-0139882 and by DOE under contract 
DE-AC05-00OR22725 with 
UT-Battelle, LLC.

\end{document}